\documentclass[aps,prd,twocolumn, floatfix, nofootinbib, superscriptaddress]{revtex4-1}
\usepackage{epsfig}
\usepackage[colorlinks,linkcolor=blue,anchorcolor=blue,citecolor=blue,urlcolor=blue,breaklinks=true]{hyperref}
\usepackage{graphics}
\usepackage{slashed}
\usepackage[dvipsnames]{xcolor}
\usepackage{amsfonts}
\usepackage{amsmath}
\usepackage{amssymb}
\usepackage{changes}

\newcommand{\sh}[1]{\slashed{#1}}

\begin{document}

\author{Chao Shi}
\email[]{shichao0820@gmail.com}
\affiliation{Physics Division, Argonne National Laboratory, Argonne, Illinois 60439, USA}

\author{C\'edric Mezrag}
\affiliation{Istituto Nazionale di Fisica Nucleare, Sezione di Roma, P.le A. Moro 2, I-00185 Roma, Italy}

\author{Hong-shi Zong}
\email[]{zonghs@nju.edu.cn}
\affiliation{Department of Physics, Nanjing University, Nanjing, Jiangsu 210093, China}
\affiliation{Joint Center for Particle, Nuclear Physics and Cosmology, Nanjing, Jiangsu 210093, China}

\title{Pion and kaon valence quark distribution functions from Dyson-Schwinger equations}
\begin{abstract}

Using realistic quark propagators and meson Bethe-Salpeter amplitudes based on the Dyson-Schwinger equations, we calculate the  pion and kaon's valence parton distribution functions (PDF) through the modified impulse approximation. The PDFs we obtained at hadronic scale have the purely valence characteristic and exhibit both dynamical chiral symmetry breaking and $SU(3)$ flavor symmetry breaking effects. A new calculation technique is introduced to determine the valence PDFs with precision. Through NLO DGLAP evolution,  our result is compared with pion and kaon valence PDF data at experimental scale. Good agreement is found in the case of pion, while deviation emerges for kaon. We point out this situation can be resolved by incorporating gluon contributions into the mesons if the pion hosts more gluons than kaon nonperturbatively.
\bigskip



\end{abstract}
\maketitle

\section{Introduction}
\label{sec:intro}
The light pseudo-scalar mesons, i.e., the pion and the kaon, play an important role in quantum chromodynamics (QCD). They are composite particles with the simplest valence content, i.e., quark and anti-quark pair, and meanwhile are the Goldstone bosons of QCD's dynamical chiral symmetry breaking (DCSB) \cite{Maris:1997hd}. Consequently, while at first sight the description of these meson could have been thought to be simple, their dual nature ensures the failure of any naive descriptions.
The determination of their structure from QCD has thus been a challenge to theory studies. Among them, the parton distribution functions (PDF) are of particular interest by revealing hadrons' parton structure and providing the non-perturbative part in the description of hard inclusive processes. In addition, the pion PDF  provides  an explanation to the up/down sea quark flavor asymmetry in the nucleon PDF through the pion clouds \cite{Thomas:1983fh}.

The Drell-Yan process has been the primary source for experimental information on pion and kaon PDF, by providing data in the valence region $1>x>0.2$. Based on the well studied  nucleon PDF, both the pion and kaon PDFs can be extracted \cite{Badier:1983mj,Conway:1989fs}. On top of this, in the low $x$ sea region $x<0.01$, HERA provided information through the deep inelastic scattering from the virtual pion cloud of the proton \cite{Adloff:1998yg,Chekanov:2002pf}. Such techniques allows to recover the internal structure of on-shell particles from scattering of off-shell ones \cite{Qin:2017lcd}.  In the future, this gap may be bridged by the tagged DIS experiment at the upgraded Jefferson Laboratory (JLab 12) \cite{Adika2015}, and possibly kaon as well.   

On the theory side, various studies, e.g., the NJL model \cite{Bentz:1999gx,RuizArriola:2002bp}, constituent quark model \cite{Suzuki:1997wv,Watanabe:2016lto} and DSEs studies \cite{Hecht:2000xa,Nguyen:2011jy,Chang:2014lva,Chen:2016sno}, have given a diversity of results. The lattice QCD is able to provide several low moments of the PDF \cite{Detmold:2003tm}. Now with the help of Large Momentum Effective field Theory (LaMET) \cite{Ji:2013dva}, it gains access to the $x$-dependence of PDF through the Quasi-PDFs \cite{Chen:2018fwa,Chen:2018xof}. Results have been obtained and keep improving, e.g., minimizing the finite-volume effects and enlarge the nucleon boost momentum for better precision \cite{Alexandrou:2015rja,Chen:2016utp}. However, it has been highlighted \cite{Xu:2018eii} that the violation of the PDF support property by quasi-PDF may reduce the relevance of the technique for high-$x$ studies. The competing Lattice QCD technique of pseudo-PDF \cite{Orginos:2017kos} might overcome this difficulty.

In this paper, we revisit the pion and kaon valence PDF within the Dyson-Schwinger equations (DSE) approach. We employ the modified impulse approximation, which was introduced in the sketch of pion and kaon PDF with simple algebraic model \cite{Chen:2016sno}. It modifies the handbag diagram (impulse approximation) and gives the valence picture of pion and kaon as a pair of fully dressed and bounded quark-antiquark. We will use the dressed quark propagators and meson amplitudes based on DSE and Bethe-Salpeter equation (BSE), and try to reveal the realistic valence picture of pion and kaon. Note that the DSEs well incorporates the QCD's DCSB and provides a faithful description for the pion and kaon as Goldstone bosons \cite{Maris:1997hd}. One of its recent successes is its prediction on the unimodal and broad profile of pion and kaon parton distribution amplitude, which recently gain support from lattice simulation \cite{Chang:2013pq,Shi:2014uwa,Cloet:2013tta,Zhang:2017bzy,Chen:2017gck}.

This paper is organized as follows. In Sec.~\ref{sec:formalism} we introduce the valence-quark PDF and  its formulation in DSEs using the modified impulse approximation. The parameterized meson amplitudes and quark propagators calculation elements will be recapitulated, along with a calculation technique demonstrating how to extract the point-wise accurate PDF based on the formulas. In Sec.~\ref{sec:pdfs}, we show our pion and kaon valence-quark PDF at hadronic scale. These PDFs are then evolved to higher scale and compared with experiment analysis. Finally, we summarize our results and give our conclusions  in Sec.~\ref{sec:sum}.

\section{Meson quark distribution in the DSE formalism.}
\label{sec:formalism}

For quark of flavor $q$ in hadron $h$, the PDF $q^{h}(x)$ is defined as the correlator 

\begin{align}
q^\pi(x)=\frac{1}{4\pi} \int d\lambda \textrm{e}^{ix P\cdot n \lambda}\langle h(P)|\bar{\psi}_q(\lambda n)\slashed{n}\psi_q(0)|h(P)\rangle_c. \label{eq:pdfdef}
\end{align}
It is the probability density for finding quark $q$ carrying the longitudinal momentum fraction $x$ of parent hadron $h$. Here the light-cone basis vector $n$ satisfies $n \cdot n=0$ and gives $P \cdot n=P^+$. The Lorentz-invariance of PDF is obvious in Eq.~(\ref{eq:pdfdef}).  

 The calculation of Eq.~(\ref{eq:pdfdef}) reduces to summing up a selection of relevant diagrams, i.e., implementing appropriate truncation scheme. Here we employ the modified impulse approximation. For $\pi^+$ it reads \cite{Chang:2014lva}  
\begin{align}
\label{eq:fcov}
u^\pi(x)&=-\textrm{Tr} \int \frac{dk^4}{2(2 \pi)^4} \delta_n^x(k_\eta)\nonumber \\
& \times  n \cdot \partial_{k_\eta} \left [ \bar{\Gamma}_\pi(k; -P) S_u(k_\eta)\right ]  \Gamma_\pi(k;P) S_d(k_{\bar{\eta}}),
\end{align}
with  $\delta_n^x(k_\eta)=\delta(n \cdot k_\eta-x n\cdot P) $. The trace should be taken in the color and Dirac space and the derivative acts only on the bracketed terms. Note that we formulated the DSEs in the Euclidean space: $P$ is the pion four momentum and $P^2=-m_\pi^2$, $n \cdot P=-m_\pi$. The quark momentum $k_\eta=k+\eta P$, $k_{\bar{\eta}}=k-(1-\eta) P$, $\eta \in [0,1]$. The final result is independent of $\eta$ due to translational invariance of the momentum integral. $S(k)$ is the dressed quark propagator  and  $\Gamma(k;P)$ is the meson Bethe-Salpeter amplitude.  In addition to the handbag diagram (impulse approximation) proportional to $\partial_{k_\eta} S(k_\eta)$, Eq.~(\ref{eq:fcov}) introduces a term proportional to $\partial_{k_\eta} \bar{\Gamma}_\pi(k;-P)$ to implement the operator insertion on the meson amplitude. This additional term respects  the nonlocal structure of the pion wave function and completes the  pion's valence picture. In the case of $K^+$, we have analogously \cite{Chen:2016sno}
\begin{align}
\label{eq:fcov2}
u^K(x)&=-\textrm{Tr} \int \frac{dk^4}{2(2 \pi)^4} \delta_n^x(k_\eta)\nonumber \\
& \times  n \cdot \partial_{k_\eta} \left [ \bar{\Gamma}_K(k; -P) S_u(k_\eta)\right ]  \Gamma_K(k;P) S_s(k_{\bar{\eta}}), \\
\bar{s}^K(x)&=-\textrm{Tr} \int \frac{dk^4}{2(2 \pi)^4} \delta_n^x(k_{\bar{\eta}})\nonumber \\
& \times   \bar{\Gamma}_K(k; -P) S_u(k_\eta)  n \cdot \partial_{k_{\bar{\eta}}} \left [ \Gamma_\pi(k;P) S_s(k_{\bar{\eta}}) \right ]. \label{eq:fcov3}
\end{align}

In the DSEs framework, the $S(k)$ and $\Gamma(k;P)$ are obtained as the solution to coupled quark's DS equation and meson's BS equation, based on interaction kernels respecting the Axial-vector Ward-Takahashi identity. Here we employ the $S(k)$ and $\Gamma(k;P)$ based on DCSB-improved kernel. It incorporates the DCSB dressing effect into the interaction kernel and improves upon the Rainbow-Ladder (RL) truncation\cite{Chang:2009zb,Chang:2011ei} in some aspects, e.g., it exposes a key role played by the dressed-quark anomalous chromomagnetic moment  in determining observable quantities \cite{Chang:2009zb} and clarifies a causal connection between DCSB and the mass splitting between vector and axial-vector mesons \cite{Chang:2011ei}. It also provides a more faithful description to the pion and kaon in terms of their PDAs \cite{Chang:2013pq,Shi:2014uwa}. However it should be noted that, Eqs.~(\ref{eq:fcov},\ref{eq:fcov2},\ref{eq:fcov3}) in principle only follows the RL truncation, i.e., the normalization condition (quark number sum rule) can be preserved automatically only with RL pion and kaon. Nevertheless, we assume in our case they provide the dominant contribution to the valence quark PDF, and the uncertainty introduced in this step doesn't exceed the generic accuracy in a valence picture description of mesons.

Solutions of the DSE-BSE with the so-called DCSB-improved kernel are available within the literature, both for the pion and the kaon \cite{Chang:2013pq,Shi:2014uwa}. In this work, we employ these results and their available parameterization, that we remind to the reader and slightly modify. The quark propagator $S(k)$ is written as the sum of two pairs of complex conjugate poles:
\begin{align}
\label{eq:spara}
S(k)=\sum_{i=1}^{2}\left [ \frac{z_i}{i \sh{k}+m_i}+\frac{z^*_i}{i \sh{k}+m^*_i} \right ],
\end{align}
with the parameter value listed in Table.~\ref{Table:parameters}. The BS amplitude $\Gamma(k;P)$, which generally takes the form 
\begin{align}
\label{eq:gammapara}
\Gamma_\pi(k;P)=&\gamma_5\biggl [ i E(k;P)+\sh{P} F(k;P)\nonumber \\
&+\sh{k} G(k;P)+[\sh{P},\sh{q}]H(k;P)\biggl ],
\end{align}
is here restricted to its dominant terms $E(k;P)$ and $F(k;P)$, which are parameterized as ($\eta=1/2$) \cite{Chang:2013pq,Shi:2014uwa}:
\begin{align}
\label{eq:fpara}
{\cal F}(k;P)&=\int_{-1}^1 d\alpha \rho_{i}(\alpha)\bigg[\frac{U_1 \Lambda^{2 n_1}}{(k^2+\alpha k\cdot P+\Lambda^2)^{n_1}}\nonumber \\ 
&+\frac{U_2 \Lambda^{2 n_2}}{(k^2+\alpha k\cdot P+\Lambda^2)^{n_2}}\bigg]\nonumber\\ 
&+\int_{-1}^1 d\alpha \rho_u(\alpha)\frac{U_3 \Lambda^{2 n_3}}{(k^2+\alpha k\cdot P+\Lambda^2)^{n_3}},\\
\rho_i(\alpha)&=\frac{1}{\sqrt{\pi}}\frac{\Gamma(3/2)}{\Gamma(1)}[C_0^{(1/2)}(\alpha)+\sigma^i_1 C_1^{(1/2)}(\alpha)\nonumber\\
&+\sigma^i_2 C_2^{(1/2)}(\alpha)] \label{eq:rho}, 
\end{align}
where $\rho_u(\alpha)=\frac{3}{4}(1-\alpha^2)$ and  \{$C_n^{(1/2)}, n=0,1,...,\infty$\} are the Gegenbauer polynomials of order $1/2$. The value of the  parameters are listed in Table.~\ref{Table:parameters}. This parameterization basically follows that in \cite{Chang:2013pq} and \cite{Shi:2014uwa}. The only difference is that herein we test with polynomial form for the weight function $\rho_\sigma(z)$. We find it modifies the end point behavior of PDF $q(x)$, but generally brings minor changes in the other regions. 

\begin{table}[t]
\caption{Representation parameters. \emph{Upper panel}: Eq.~(\ref{eq:spara}) -- the pair $(x,y)$ represents the complex number $x+ i y$.  \emph{Lower panel}: Eqs.~(\ref{eq:gammapara},\ref{eq:fpara},\ref{eq:rho}).  (Dimensioned quantities in GeV).
\label{Table:parameters}
}
\begin{center}
\begin{tabular*}
{\hsize}
{
@{\extracolsep{0ptplus1fil}}
c@{\extracolsep{0ptplus1fil}}
c@{\extracolsep{0ptplus1fil}}
c@{\extracolsep{0ptplus1fil}}
c@{\extracolsep{0ptplus1fil}}
c@{\extracolsep{0ptplus1fil}}}\hline
  & $z_1$ & $m_1$  & $z_2$ & $m_2$ \\
$u$ &    $(0.44,0.28)$ & $(0.46,0.18)$ & $(0.12,0)$ & $(-1.31,-0.75)$ \\
$s$ &    $(0.43,0.30)$ & $(0.55,0.22)$ & $(0.12,0.11)$ & $(-0.83,0.42)$ \\
\hline
\end{tabular*}

\begin{tabular*}
{\hsize}
{
l@{\extracolsep{0ptplus1fil}}
l@{\extracolsep{0ptplus1fil}}
c@{\extracolsep{0ptplus1fil}}
c@{\extracolsep{0ptplus1fil}}
c@{\extracolsep{0ptplus1fil}}
c@{\extracolsep{0ptplus1fil}}
c@{\extracolsep{0ptplus1fil}}
c@{\extracolsep{0ptplus1fil}}
c@{\extracolsep{0ptplus1fil}}
c@{\extracolsep{0ptplus1fil}}}\hline
   & $U_1$ & $U_2$ & $U_3$ &$n_1$ &$n_2$ &$n_3$ & $\sigma^i_1$ & $\sigma^i_2$ & $\Lambda$ \\\hline
 E$_\pi$ & $2.76$ & $-1.84$ & $0.04$ & $4$
    & $5$ & $1$&0.0 &2.2&1.41 \\
   F$_\pi$ & $1.46$ & $-0.97$ & $0.006$ &$4$
    & $5$ & $1$&0.0&-0.5& 1.13   \\
 E$_K$ & $2.98$ & $-2.0$ & $0.025$ & $4$
    & $5$ & $1$&-0.4 &1.0&1.35 \\
   F$_K$ & $0.86$ & $-0.30$ & $0.004$ &$4$
    & $6$ & $1$&-0.4&-1.0& 1.20   \\\hline
\end{tabular*}
\end{center}
\vspace*{-4ex}

\end{table}

Now we can calculate the pion and kaon PDFs. The starting point is to look at their moments 
\begin{align}
\langle x^m \rangle=\int_0^1 dx x^m q(x).
\end{align}
Conventionally, one can calculate many moments, for instance 50, and try to reconstruct the original function $q(x)$ \cite{Chen:2016sno}. In practice however, this requires good intuition and guess on its analytic form. The deviation between the original function and the guessed analytic form brings ambiguity to the reconstruction. In this connection, we employ a method that could determine $q(x)$ point-wisely \cite{Mezrag:2014tva,Mezrag:2016hnp}, as we explain below.

We still start with the moments $\langle x^m \rangle$, i.e., for pion
\begin{align}
\langle x^m \rangle&=-\textrm{Tr} \int \frac{dk^4}{2(2 \pi)^4} \frac{1}{|P \cdot n|}\bigg(\frac{k_\eta \cdot n}{P \cdot n}\bigg)^m\nonumber \\
& \times  n \cdot \partial_{k_\eta} \left [ \bar{\Gamma}_\pi(k; -P) S_u(k_\eta)\right ]  \Gamma_\pi(k;P) S_d(k_{\bar{\eta}}). 
\end{align}  
Using the Feynman parameterization technique, the loop momentum integral is replaced by integration over three independent Feynman parameters, i.e., $x_1$, $x_2$, $x_3$. Adding up the two integral variables from weight function $\rho(\alpha)$ in Eq.~(\ref{eq:rho}), i.e., $\alpha_1$ and $\alpha_2$, we are left with a 5-dimension integral. 
\begin{align}
\label{eq:mompre}
\langle x^m \rangle=\prod_{i=1}^3 \int_0^1 dx_i \prod_{j=1}^2\int_{-1}^1  d\alpha_j H(x_i, \alpha_j,m).
\end{align}
Here $H$ is some function to be integrated, with $m$ one of its variables. The idea is then to perform a transform of integral variables to rewrite the right hand side of Eq.~(\ref{eq:mompre}) as
\begin{align}
\label{eq:momfinal}
\langle x^m \rangle= \int_0^1 dx' x'^m \prod_{i=1}^4\int  dx'_i G(x', x'_i),
\end{align}
which is feasible as we show in the Appendix~\ref{sec:App}. The $G(x', x'_i)$ must no longer  depend on $m$. One then quickly identifies that $f(x)=\prod_{i=1}^4\int  dx'_i G(x, x'_i)$, which can be computed numerically. 

\section{pion and kaon valence quark distribution functions.}
\label{sec:pdfs}
\begin{figure}[htbp]
\includegraphics[width=9cm]{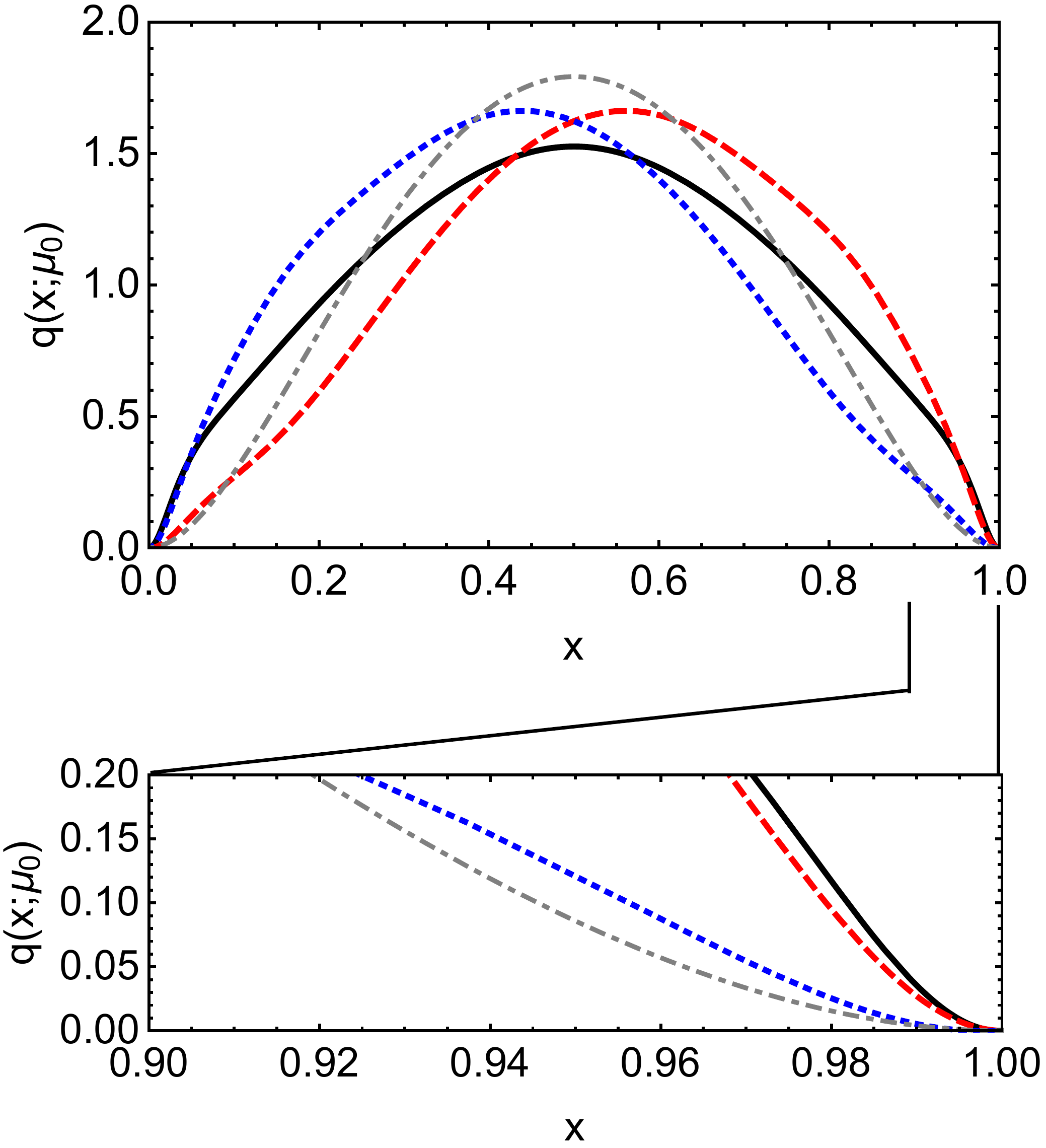}
\caption{\label{fig:pdf0}The pion and kaon PDF at $\mu_0=520$ MeV. The solid black line depict  $u^{\pi}(x,\mu_0)$ from our calculation, with blue dotted curve for $u^K(x,\mu_0)$ and red dashed curve for $\bar{s}^K(x,\mu_0)$. The gray dot-dashed curve is obtained with algebraic model in \cite{Chang:2014lva}. The lower plot zooms into the large $x$ region of upper plot.}
\end{figure}

The valence quark distributions of pion and kaon are plotted in Fig.~\ref{fig:pdf0}. These PDFs are at certain low hadronic scale $\mu_0$ where all the  sea quarks and gluons are absorbed into the dressed quarks. We expect the natural scale at which this picture is a good approximation to be low, typically of the size or below the nucleon mass. The value of $\mu_0$ will be estimated later. 

Let's first look at the pion $u^\pi(x;\mu_0)$. The distribution is symmetric with respect to $x=1/2$, in line with the $u-d$ isospin symmetry. In this case, the quark and antiquark  each carries half of the meson's light front momentum automatically, i.e., $\langle x \rangle^\pi_u=0.5$. Note that the quark number sum rule has been implemented $\langle x^0 \rangle^\pi_u=1$ as normalization condition. If we take only the handbag diagram, then $\langle x^0 \rangle^\pi_u$ continues to be one, model independently \cite{Chang:2014lva}, but  $\langle x \rangle^\pi_u$ becomes $0.46$. The modification term proportional to $\partial_{k_\eta} \bar{\Gamma}$  therefore collects  the 8\% missing  momentum fraction back to the dressed quarks. This valence picture is further confirmed in the case of kaon, i.e., $\langle x\rangle^K_u+\langle x \rangle^K_{\bar{s}}=1$, more specifically $u^K(x)=\bar{s}^K(1-x)$, which is the consequence of  momentum conservation  in terms of dressed quarks. 

 Another observation is that the solid curve is broader than the dot-dashed one. The latter shows the PDF computed from an algebraic model of the propagators and Bethe-Salpeter amplitudes developed in Ref.\cite{Chang:2013pq} and adapted by the authors of Ref.\cite{Chen:2016sno} (see, e.g., Eq.(8) of \cite{Chen:2016sno}). Such algebraic models, based on simple but insightful parameterizations of $S(k)$ and $\Gamma_\pi(k)$ have yielded interesting results and discussions both in the meson sector (see e.g. \cite{Gao:2014bca,Mezrag:2016hnp,Chouika:2017rzs}) and in the baryon one \cite{Mezrag:2017znp}. In the present case, the realistic quark propagator and BS amplitudes give a broader $u^\pi(x;\mu_0)$ than the one obtain with the algebraic model.
 The broadness discrepancy can be quantified by studying the $\langle(2x-1)^2\rangle_u^\pi$ of the distributions $u^\pi(x;\mu_0)$. 
The present computation of the PDF yields $\langle(2x-1)^2\rangle_u^\pi=0.20$ while the algebraic model of Ref.\cite{Chen:2016sno} gives $\langle(2x-1)^2\rangle_u^\pi=0.15$. The situation is similar with kaon, i.e., $\langle(2x-1)^2\rangle_u^K=0.176$ versus $\langle(2x-1)^2\rangle_u^K=0.134$ obtained with the algebraic model of Ref.\cite{Chen:2016sno}. From our perspective, this difference traces back to a faithful representation of the DCSB effect: the $S(k)$ and $\Gamma(k;P)$ encoding  realistic DCSB effect typically generate a parton distribution amplitude (PDA) which is again broader than the one coming from the algebraic model. In the present case, the DB-kernel pion generates a pion PDA $\phi(x)\approx 1.81(x(1-x))^{0.31}(1-0.12 C_2^{0.81}(2x-1))$, which is broader than $\phi(x)=6x(1-x)$ from algebraic model \cite{Chang:2013pq}. This feature is reflected in the case of our PDF.

 The underlying connection between PDA and PDF can be viewed from the perspective of pion's leading twist light front wave function $\psi(x,\bold{k}_\perp^2)$ \cite{Burkardt:2002uc,Braun:1989iv}. The PDA is defined as \cite{Lepage:1980fj}
\begin{align}
\phi(x,\mu^2)&=\int^{\mu^2} d^2 \bold{k}_\perp \psi(x,\bold{k}_\perp^2),
\end{align}
while
\begin{align}
q(x,\mu_0^2)&\approx \int d^2 \bold{k}_\perp |\psi(x,\bold{k}_\perp^2)|^2
\end{align} 
approximates the PDF at some low hadronic scale in the absence of higher Fock state \cite{Burkardt:2002uc} and neglecting the higher twist wave function \cite{Burkardt:2002uc,Mezrag:2016hnp}. The PDA and PDF are therefore implicitly related and the broadness in $\phi(x,\mu^2)$ would be reflected in $q(x,\mu_0^2)$, as we have observed above.

Fig.~\ref{fig:pdf0} also shows the SU(3) flavor symmetry breaking in the kaon PDF, as $u^K(x) \ne \bar{s}^K(x)$. The heavier $s$ quark carries a larger fraction of the meson momentum, i.e., $\langle x \rangle ^K_{\bar{s}}=0.55$, with the rest $45\% $ carried by $u$ quark. We remind that for the kaon PDA, a similar result is obtained, i.e., $\langle(2x-1)\rangle_\phi=0.04$ \cite{Shi:2014uwa}. The 10\% difference in kaon's $s$ and $u$ quark PDF is significantly smaller than their current quark mass ratio $m_s/m_u \gtrsim 20$. Therefore the SU(3) flavor symmetry is strongly masked by the dressing  effect on light quarks through DCSB.

The end point behavior of the valence PDFs is shown in the lower plot in Fig.~\ref{fig:pdf0}. Theoretically, as $x\rightarrow 1$, one quark carries almost all the plus momentum of its parent meson and gets far off shell. Then the pQCD becomes valid and predicts the power behavior of valence PDF $\sim (1-x)^2$ as $x \rightarrow 1$ \cite{Ji:2004hz}. This power behavior is respected by all our curves as shown by the plot. It starts from some inflection point around $x\gtrsim 0.95$, signaling the transition from soft nonperturbative QCD dynamics to hard pQCD interactions.

We futher parameterize our PDFs with 
\begin{align}
\label{eq:paraPDF}
q(x;\mu_0)=30[x(1-x)]^2\left[1+\sum_{j=1}^{j_m}a_j C_j^{5/2}(2x-1)\right].
\end{align}
We find with $j_m=10$ the curves can well be represented by parameters in Table.~\ref{tab:pdfparas}. 

\begin{table}[tb]
\centering
\caption{Fitting parameters in Eq.~(\ref{eq:paraPDF}) for $u^\pi(x,\mu_0)$ and $\bar{s}^K(x,\mu_0)$.   \label{tab:pdfparas}}
\begin{tabular*}
{\hsize}
{
c@{\extracolsep{0ptplus1fil}}
r@{\extracolsep{0ptplus1fil}}
r@{\extracolsep{0ptplus1fil}}
r@{\extracolsep{0ptplus1fil}}
c@{\extracolsep{0ptplus1fil}}
c@{\extracolsep{0ptplus1fil}}
l@{\extracolsep{0ptplus1fil}}
l@{\extracolsep{0ptplus1fil}}}
\hline
 & $a_1$ & $a_2$ & $a_3$ & $a_4$&  $a_5$ & \\\hline
$\pi$&0&0.125&0&0.0463&0\\
K&0.137&0.0894&0.0313&0.0292&0.00671\\\hline\hline
 & $a_6$ & $a_7$ & $a_8$ & $a_9$&  $a_{10}$ & \\\hline
$\pi$&0.0181&0&0.00651&0&0.00152\\
K&0.00801&0.00178&0.00214&0.000728&0.000518\\\hline
\end{tabular*}
\end{table}

\begin{figure}[htbp]
\includegraphics[width=9cm]{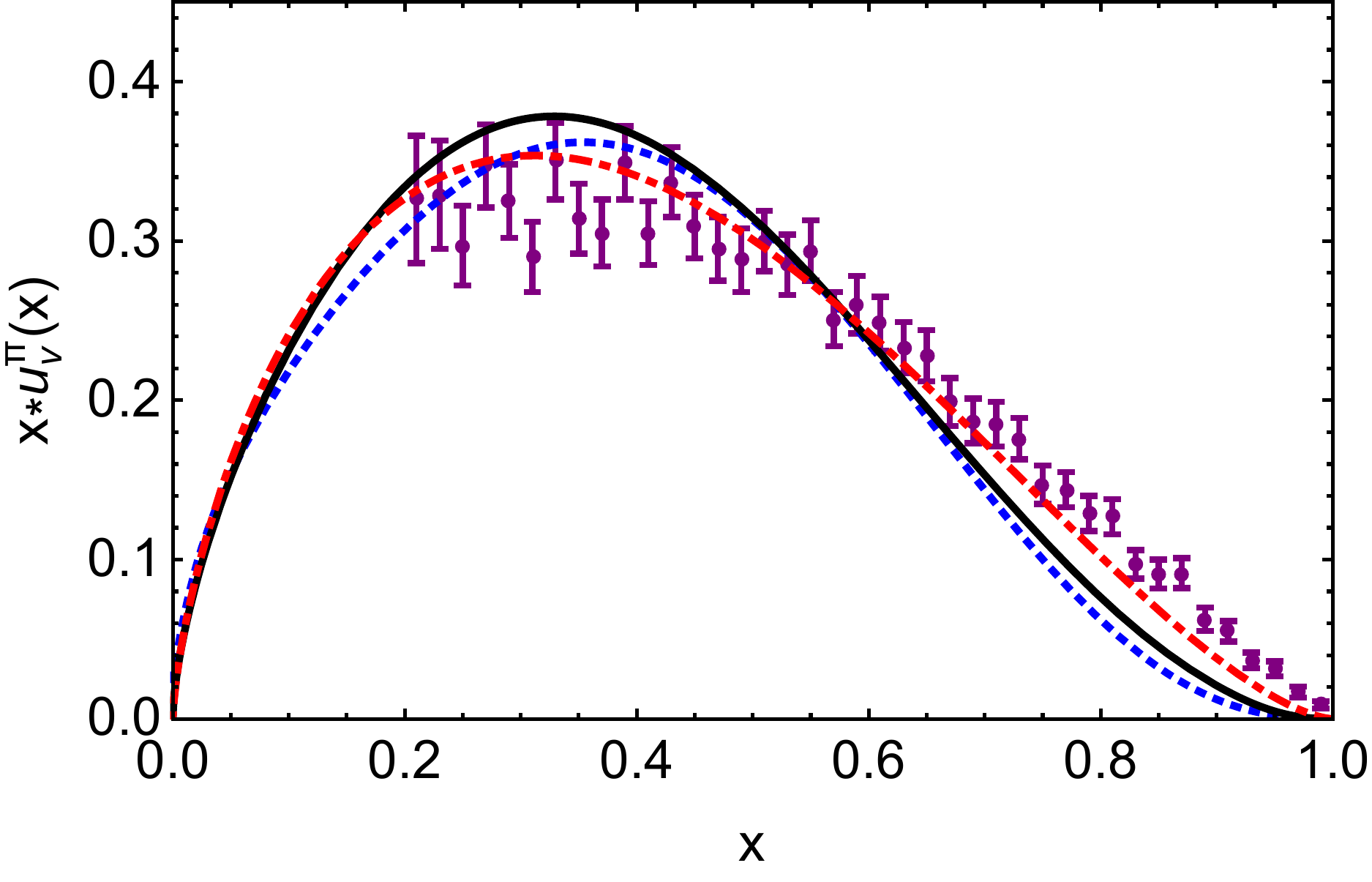}
\caption{\label{fig:pionpdf1}Our pion valence PDF NLO evolved to $\mu_4=4$ GeV is displayed as the black solid curve. NLO analysis of Fermilab E-615 pionic Drell-Yan data with soft-gluon resummation  \cite{Aicher:2010cb}  is plotted as blue dotted curve ($\mu=4$ GeV). Without soft-gluon resummation, NLO analysis gives the red dot-dashed line ($\langle M_\gamma \rangle=5.2$ GeV) \cite{Wijesooriya:2005ir}. The purple filled cricles are LO analysis result ($\langle M_\gamma \rangle=5.2$ GeV) \cite{Conway:1989fs}. }
\end{figure}

We then perform the NLO DGLAP evolution on valence-quark distribution $u^\pi_v(x)=u^\pi(x)-\bar{u}^\pi(x)$ using the QCDNUM package \cite{Botje:2010ay}.  The strong coupling constant is set to be the optimal value in NLO global PDF analysis $\alpha_s(1 \textrm{GeV})=0.491$ \cite{Martin:2009iq} and the variable flavor number scheme (VFNS) is taken. 
It is  found that for $u_v^\pi(x,\mu_0)$, $\mu_0=520$ MeV produces $\langle x \rangle_v^\pi \sim 0.24$ at $\mu_2= 2 \textrm{GeV}$, close to  the $\pi N$ Drell-Yan data analysis $2\langle x \rangle_v^{\mu_2}=0.47(2)$ \cite{Sutton:1991ay,Gluck:1999xe} and lattice simulation  $2\langle x \rangle_v^{\mu_2}=0.48(4)$ \cite{Detmold:2003tm}. The evolved valence quark distribution $u^\pi_v(x,\mu_4)$ with $\mu_4=4$ GeV is plotted in Fig.~\ref{fig:pionpdf1}. As it can be seen, our result generally agrees with existing data analysis. Especially it favors the result from \cite{Aicher:2010cb} when $x\gtrsim 0.6$. In this connection, LO analysis found almost linear decrease at large x, i.e., $\sim(1-x)^{2+\beta}$ with $\beta=-0.74$ \cite{Conway:1989fs} while NLO analysis almost halved this value, i.e., $\beta=-0.40$ \cite{Aicher:2010cb}. The authors of \cite{Aicher:2010cb} show that the logarithmic threshold resummation brings considerable reduction at large $x$, i.e., $\beta=0.34$ and therefore agrees with pQCD prediction $\beta >0$. Since we already have  $\beta=0$ at $\mu_0$, DGLAP evolution to higher scale further shifts the support of $u_v^\pi(x;\mu_0)$ from larger $x$ to smaller $x$ and therefore increases the value of $\beta$.

\begin{figure}[htbp]
\includegraphics[width=9cm]{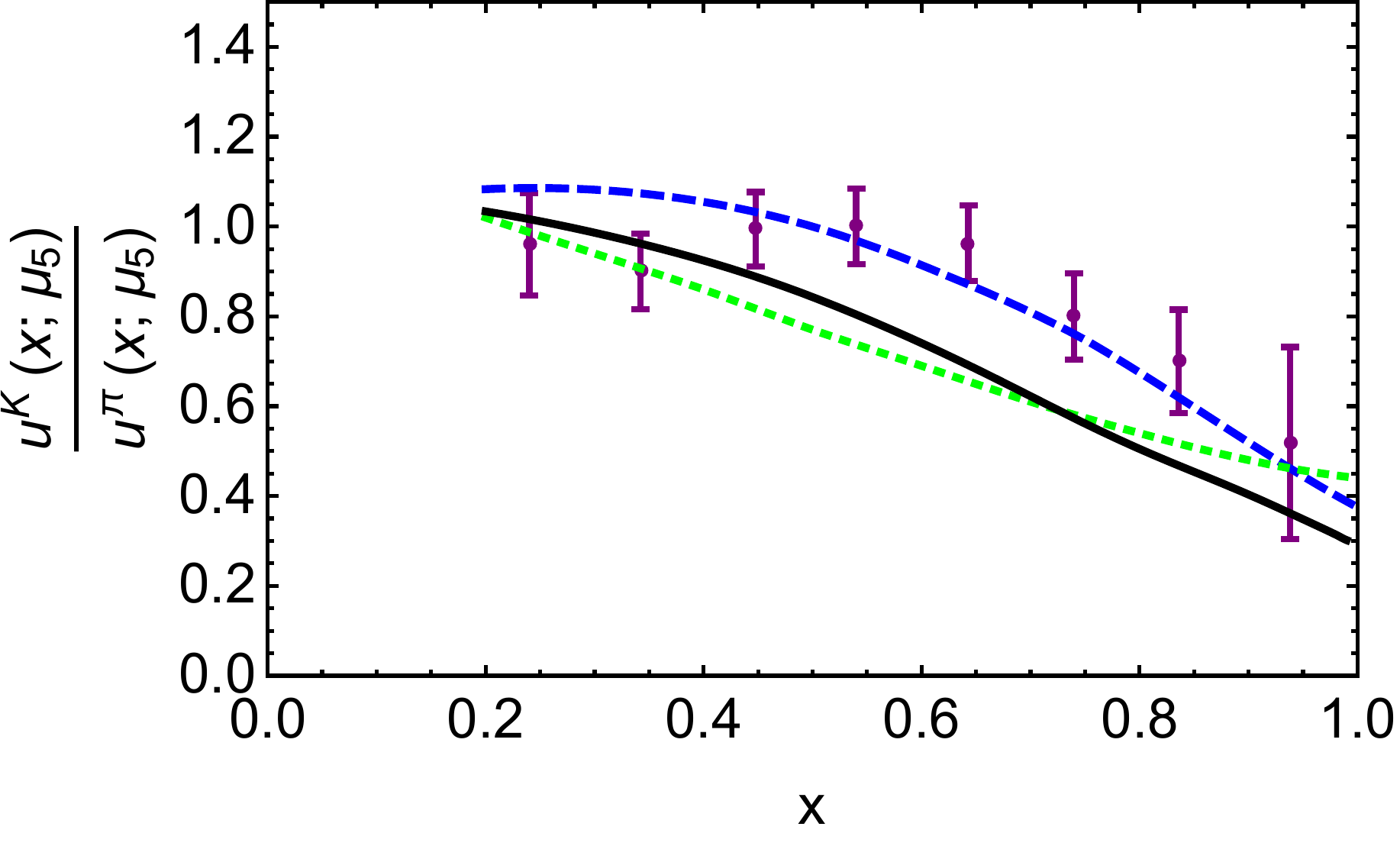}
\caption{\label{fig:kaonpdf1}The ratio of $\bar{u}$ distribution function in $K^-$ to that of $\pi^-$, i.e., $\bar{u}^{K}(x;\mu_5)/\bar{u}^{\pi}(x;\mu_5)$ with $\mu_5=5.2$ GeV. The blue filled circles are experimental data from \cite{Badier:1980jq}. Our result is depicted as the black solid curve. Green dotted curve is the NJL model calculation with proper-time regularization \cite{Hutauruk:2016sug} and blue dashed curve is from \cite{Chen:2016sno}.}
\end{figure}

We finally depict the the ratio of $\bar{u}$ distribution function in $K^-$ to that of $\pi^-$, i.e., $\bar{u}^{K}(x;\mu_5)/\bar{u}^{\pi}(x;\mu_5)$ with $\mu_5=5.2$ GeV, in Fig.~\ref{fig:kaonpdf1}. Generally speaking, our result undershoots many data points in the valence region, but shows agreement at low and large $x$ region. We point out same situation occurs in the NJL model calculation with proper-time regularization (green dotted curve) \cite{Hutauruk:2016sug}, which also starts with the valence picture for mesons. The cause of the deviation at intermediate $x$ region deserves further investigation but one clue can be found in  \cite{Chang:2014lva,Chen:2016sno}. Therein the authors point out that nonperturbatively, the pion should have more gluon content than the kaon does at certain hadronic scale $\mu_0'$. By incorporating the gluon effect, they found good agreement with valence PDF data for both the pion and the kaon. Herein if we incorporate this gluon effect into our  $\bar{u}^{\pi/K}(x;\mu_0)$ and bring them to $\bar{u}^{\pi/K}(x;\mu_0')$ \footnote{In terms of BSE, this means we need to consider an additional component of mesons, i.e., $q\bar{q}g$.}, then the $\bar{u}^{\pi}(x;\mu_0)$ and $\bar{u}^{K}(x;\mu_0)$ would both shift to lower $x$ but $\bar{u}^{\pi}(x;\mu_0)$ would shift more, since more of the quark momentum in the pion should be carried away by gluons than in the kaon. The consequence is that  $\bar{u}^{K}(x;\mu_0')$ would get close to $\bar{u}^{\pi}(x;\mu_0')$. Note that in the large $x$ region the gluon effect is suppressed \cite{Chang:2014lva} and  the ratio changes little, i.e., $\bar{u}^{K}(x;\mu_0')/\bar{u}^{\pi}(x;\mu_0')|_{x\rightarrow 1}=\bar{u}^{K}(x;\mu_0)/\bar{u}^{\pi}(x;\mu_0)|_{x\rightarrow 1}$. In this way, the overall outcome  would be raising the ratio  from $\bar{u}^{K}(x;\mu_0)/\bar{u}^{\pi}(x;\mu_0)$ to $\bar{u}^{K}(x;\mu_0')/\bar{u}^{\pi}(x;\mu_0')$  in the intermediate $x$ region with the large $x$ region unchanged. After DGLAP evolution, the curve $\bar{u}^{K}(x;\mu_5)/\bar{u}^{\pi}(x;\mu_5)$ should inherit this remedy and get closer to the data.

\section{summary}
\label{sec:sum}
Starting with the solution of the Bethe-Salpeter equation for the pion, in a beyond rainbow-ladder truncation of QCD’s Dyson-Schwinger equations, we have computed the valence-quark PDF of pion and kaon within the modified impulse approximation.  These PDFs give the purely valence picture of pion and kaon at hadronic scale, and exhibit many properties stemming from QCD. For instance, the dynamical chiral symmetry breaking generally broadens the PDFs at hadronic scale, similarly to the case of parton distribution amplitude. The $SU(3)$ flavor symmetry breaking, masked by the DCSB, causes typically $10\%$ asymmetry in the kaon's PDFs. At large $x$, all our PDFs  decreases with the power behavior $\sim (1-x)^2$, respecting the pQCD prediction. We then evolve these valence quark distribution functions to experimental scale. Despite good agreement with the pion valence PDF, the ratio $\bar{u}^{K}(x;\mu_5)/\bar{u}^{\pi}(x;\mu_5)$ generally undershoots the data. 
We therefore sketch a resolution to this discrepancy, based on the argument that the pion hosts more gluons than the kaon at hadronic scale \cite{Chen:2016sno}. Nevertheless in the DSE-BSE framework, a conclusive verification of this problem calls for a nonperturbative study on the pion and kaon bound state equations incorporating $q\bar{q}g$ component.

\bigskip

\appendix
\section{}
\label{sec:App}
In obtaining Eq.~(\ref{eq:momfinal}) from Eqs.~(\ref{eq:mompre}), we find the following integral variable transformation useful,
\begin{align}
\prod_{i=1}^3 \int_0^1 dx_i \prod_{j=1}^2\int_{-1}^1  d\alpha_j= \prod_{i=1}^5 \int_0^1 da_i |J_a|+ \prod_{i=1}^5 \int_0^1 db_i |J_b|.
\end{align}
$|J_a|$ and $|J_b|$ are the Jacobian determinants of the new integral variables $a_i$ and $b_i$. We introduce some auxiliary variables as
\begin{align}
u&=(1-\alpha_1)/2\\
v&=(2-x_3-x_3 \alpha_2)(1-x_2)/2\\
z&=u x_1+v(1-x_1),
\end{align}
then the variable transformation is
\begin{align}
a_1&=z \\
a_2&=\frac{u}{z}\\
a_3&=\frac{x_1 (1-u)}{1-z}\\
a_4&=\frac{x_2}{1-v}\\
a_5&=\frac{x_3-(1-v/(1-x_2))}{v/(1-x_2)}
\end{align}
and $b_i=a_i$ except $b_2=(u-z)/(1-z)$ and $b_3=x_1 u/z$. In practice, this brings Eq.~(\ref{eq:mompre}) to the form 
\begin{align}
\langle x^m \rangle= \int_0^1 da_1 a_1^m (G_0+G_1 m+G_2 m^2+G_3 m^3)
\end{align}
where $G_i$'s are functions of $a_1$ and has no dependence in $m$. To further reduce it to Eq.~(\ref{eq:momfinal}), we need to remove terms proportional to $m^i$ with $i>0$. We exemplify with the term $ \int_0^1 da_1 a_1^m m^2 G_2 $. The starting point is the identity
\begin{align}
\int_0^1 da_1 \frac{\textrm{d}^2(a_1^{m+2} G_2)}{\textrm{d}a_1^2}=0.
\end{align}
The equality can be checked numerically. Fully expand the derivative in the integrand and one gets 
\begin{align}
\int_0^1 da_1 a_1^m m^2 G_2&=\int_0^1 d a_1 a_1^m[-m(3 G_2+2a_1 G_2')\nonumber\\
&-(2 G_2+4a_1 G_2'+a_1^2 G_2'')].
\end{align}
The term proportional to $m^2$ is reduced to terms of $m^1$ and $m^0$. Therefore in practice we start from the $m^3$ term and employing similar procedures iteratively until all the terms proportional to $m^i$ with $i>0$ are removed, leaving only Eq.~(\ref{eq:momfinal}).

\acknowledgments

The authors would like to thank Ian Clo\"et, Craig Roberts and Peter Tandy for beneficial discussions.
This work was supported by the U.S. Department of Energy, Office of Science, Office of Nuclear Physics, contract no. DE-AC02-06CH11357; and the Laboratory Directed Research and Development (LDRD) funding from Argonne National Laboratory, project no. 2016-098-N0 and project no. 2017-058-N0; National Natural Science Foundation of China (11475085, 11535005, 11690030) and National Major state Basic Research and Development of China (2016YFE0129300).

\bibliography{PiKvaPDF}

\end{document}